\documentclass{IEEEtran}
\usepackage{amsmath,amsfonts}
\usepackage{algorithmic}
\usepackage{algorithm}
\usepackage{array}
\usepackage[caption=false,font=normalsize,labelfont=sf,textfont=sf]{subfig}
\usepackage{textcomp}
\usepackage{stfloats}
\usepackage{url}
\usepackage{verbatim}
\usepackage{graphicx}
\usepackage{cite}
\usepackage{xcolor}
\usepackage{balance}
\usepackage{amsmath}
\usepackage{booktabs,mathptmx,siunitx}
\usepackage{multirow}
\usepackage{cite}
\usepackage{wasysym}
\usepackage{flushend}
\usepackage{xcolor}
\usepackage{balance}
\usepackage{amsmath}
\usepackage{booktabs,mathptmx,siunitx}
\usepackage{multirow}
\usepackage{cite}
\usepackage{wasysym}
\usepackage{flushend}
\usepackage{wasysym}
\usepackage{cite}
\usepackage{amsmath,amssymb,amsfonts,bm}
\usepackage{algorithmic}
\usepackage{graphicx}
\usepackage{textcomp}
\usepackage{gensymb}

\def\BibTeX{{\rm B\kern-.05em{\sc i\kern-.025em b}\kern-.08em
    T\kern-.1667em\lower.7ex\hbox{E}\kern-.125emX}}
\begin{document}
\title{Suitability of Common Ingestible Antennas for~Multiplexed~Gastrointestinal Biosensing}
\author{Erdem Cil, \IEEEmembership{Student Member, IEEE}, Ícaro V. Soares,~\IEEEmembership{Member, IEEE}, Domitille Schanne, Ronan~Sauleau,~\IEEEmembership{Fellow,~IEEE},  and~Denys~Nikolayev,~\IEEEmembership{Senior Member, IEEE}
\thanks{Manuscript received July 5, 2024. This work was supported  in part by the French \textit{Agence Nationale de la Recherche} (ANR) under grant ANR-21-CE19-0045 (project ``MedWave''), in part by the European Union through European Regional Development Fund (ERDF), Ministry of Higher Education and Research, CNRS, Brittany region, Conseils Départementaux d’Ille-et-Vilaine and Côtes d’Armor, Rennes Métropole, and Lannion Trégor Communauté, through the CPER Project CyMoCod. \textit{(Corresponding author: Denys Nikolayev.)}}
\thanks{E. Cil is with the Univ Rennes, CNRS, IETR~-- UMR 6164, FR-35000 Rennes, France and with BodyCAP, FR-14200 Hérouville St Clair, France.}
\thanks{Í. V. Soares, D. Schanne, R. Sauleau, and D. Nikolayev are with the Univ Rennes, CNRS, IETR~-- UMR 6164, FR-35000 Rennes, France (e-mail: denys.nikolayev@deniq.com).}
}

\maketitle
\begin{abstract}
Ingestible sensor devices, which are increasingly used for internal health monitoring, rely on antennas to perform sensing functions and simultaneously to communicate with external devices. Despite the development of various ingestible antennas, there has been no comprehensive comparison of their performance as biosensors. This paper addresses this gap by examining and comparing the suitability of three common types of ingestible antennas---dipole, patch, and loop---as biosensors for distinguishing gastrointestinal tissues (stomach, small intestine, and large intestine) based on their electromagnetic properties. The antennas studied in this work conform to the inner surface of biocompatible polylactic acid capsules with varying shell thicknesses and operate in the 433 MHz Industrial, Scientific, and Medical  band. The comparison is performed in gastrointestinal tissues using several antenna parameters: 1) \textit{Sensing Capability}: Changes in the phase of the reflection coefficient in the tissues are selected as the sensing parameter. 2) \textit{Robustness}: The frequency interval ($\Delta f_{i}$) in which the antennas are matched ($|S_{11}| <$ --10~dB) in all the tissues and the maximum change in the center frequency ($f_{c}$) in different tissues are examined. 3) \textit{Radiation Performance}: The gain and radiation efficiency of the antennas are examined. The effect of shell thickness on gain and radiation efficiency at 434 MHz is presented. Additionally, the radiation efficiency at various frequencies allocated for medical communications is compared with the theoretical maximum achievable efficiencies. These comprehensive data provide valuable information for making engineering decisions when designing multiplexed biosensor antennas for ingestible applications.
\end{abstract}


\section{Introduction}
\label{sec:introduction}
\IEEEPARstart{R}{ecent} advancements in bioelectronics have enabled the development of miniaturized implantable and ingestible biosensor devices that can monitor physiological parameters inside the human \cite{5724292, sharma,9300178} or animal body\cite{6920091, 10138759, 10287266}. Due to the possibility of minimally invasive real-time monitoring of vital parameters, wireless miniaturized biosensor devices are attracting increasing attention and promise new methods to advance the field of wireless bioelectronics \cite{5597914, 7855707, katz2014implantable}. Different application fields of such devices include metabolite monitoring \cite{6374200}, hypertension monitoring \cite{9471849}, blood flow sensing \cite{6213087, 8966302}, blood glucose monitoring\cite{4956982, 8910407, 7509619}, wireless endoscopy\cite{10045709, 9743772, 9217950}, gastrointestinal (GI) pH, temperature \cite{6767151}, and gas sensing \cite{nature2}.

The antenna integrated into an ingestible biosensor device (the ingestible antenna) plays a significant role in the overall wireless performance of the device \cite{10274784}. The quality of the communication link with an off-body device depends heavily on the antenna design~\cite{6413167}. Designing ingestible antennas presents several unique challenges. They include robustness against nondeterministic properties of the tissues, material restrictions due to biocompatibility requirements, size and shape limitations due to the confined space in the capsule, interactions with other components inside the capsule (e.g. battery), and various electromagnetic (EM) loss mechanisms that affect the operation of the antenna \cite{6293992, 9464241, 8607978, 7934096}. To address these challenges, different types of antennas have been proposed in literature based on, e.g., slot \cite{8012537}, loop \cite{5643917}, patch \cite{9667256}, and bowtie \cite{9762631} configurations.

The loss mechanisms associated with the operation of in-body antennas can be divided into three main categories: near-field, propagation, and reflection losses \cite{6905770, nikolayev_electromagnetic_2018, 6025277, 8606957}. Although near-field and reflection losses can be partially mitigated using proper insulating or matching layers \cite{5617242, 9677970, 9245493}, they are significant in a realistic in-body communication link \cite{8606957}. Specifically, near-field losses occur when the reactive near-fields extend into the lossy tissue surrounding the antenna, affecting parameters such as input impedance, radiation pattern, and radiation efficiency. For ingestible antennas, the near-fields extend into GI tissues and their contents: stomach (ST), small intestine (SI), and large intestine (LI). These tissues have varying EM properties (relative permittivity $\varepsilon_r$ and electrical conductivity $\sigma$) as shown in Table \ref{tab:emproperties} for 434~MHz based on averaged data reported by Gabriel~\textit{et~al.} \cite{gabriel}. As the capsule advances through the GI tract, the ingestible antenna encounters a highly variable EM environment in its near-field zone \cite{8345685}. These EM variations during the capsule transition lead to statistically consistent relative changes in the antenna parameters in different tissues, such as detuning of its input impedance and fluctuations in its radiation efficiency~\cite{erdemlast}.

\begin{figure*}
\centering
\includegraphics[width=1.6\columnwidth]{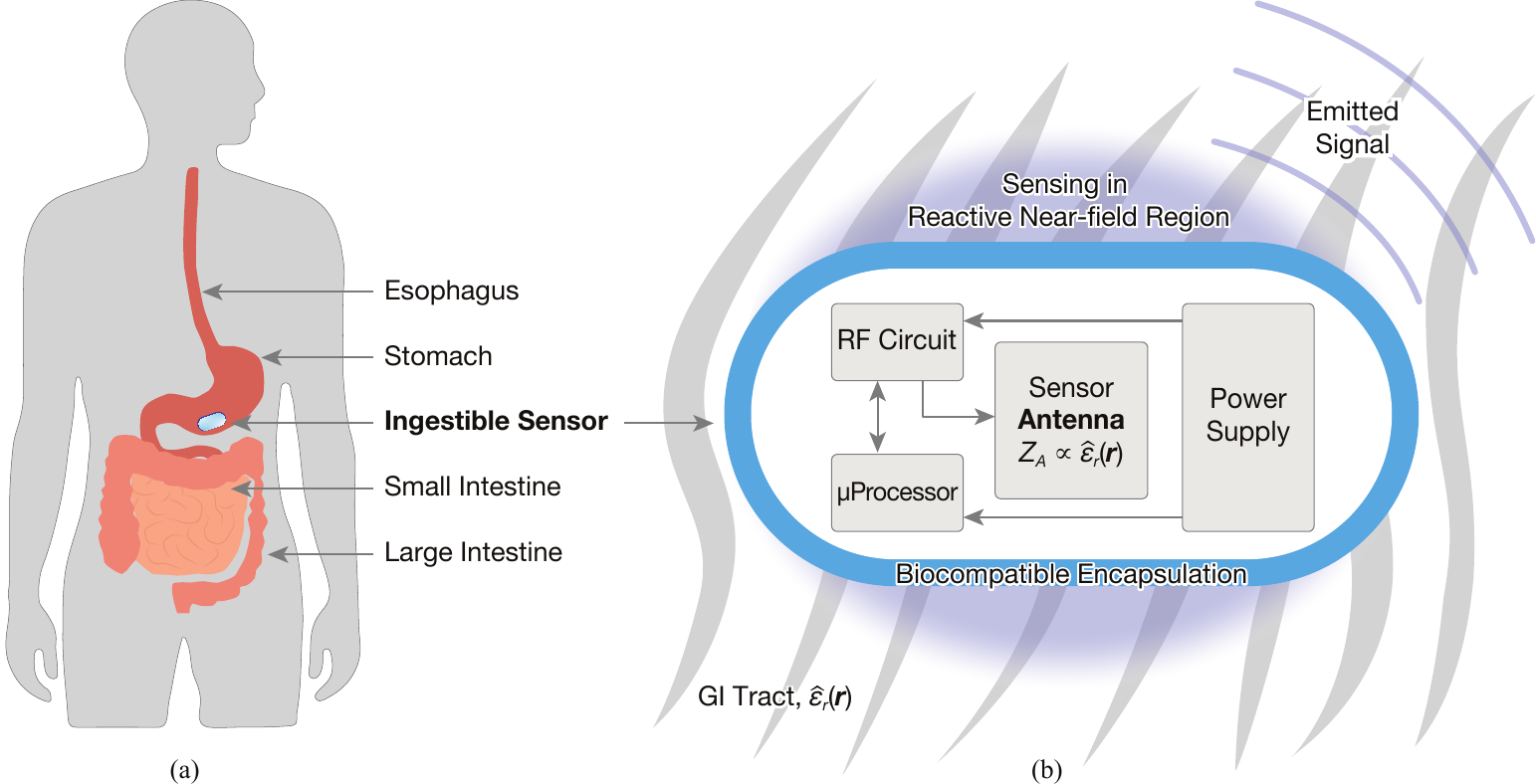}
\caption{General overview of the concept examined in this paper. (a) Human GI system with the capsule including the ingestible sensor antenna. (b) General model of the wireless biosensor capsule showing required elements to operate it and two operations realized by the capsule: 1) sensing in reactive near-field region where near-fields radiated by the antenna extend into the tissues, resulting in changes in antenna parameters in different tissues and 2) communication with the outside of the body where collected data is transmitted to an off-body device via propagation of EM-fields through the lossy tissues.}
\label{fig:concept}
\end{figure*}

In this context, this paper examines several operational parameters of three common ingestible antenna types (dipole, patch, and loop) within the GI tract to analyze and compare their suitability for GI sensing throughout ST, SI, and LI. GI sensing refers to distinguishing different GI tissues and tracking the real-time relative location of the capsule in the GI tract, which can aid in diagnosing specific pathologies such as gastroparesis~\cite{10149529}. The examination is conducted in the 433 MHz Industrial, Scientific, and Medical (ISM) band using homogeneous phantoms representing GI tissues. The concept studied in the first part of the paper is visualized in Fig.~\ref{fig:concept}. In a realistic scenario, as depicted in Fig.~\ref{fig:concept}, tissues exhibit heterogeneous and nondeterministic properties, and the GI tract contains continuously changing contents, further affecting the EM variations encountered by the antenna. Previous proof-of-concept studies conducted by the authors with an \textit{ex vivo} measurement setup demonstrated that the results obtained with the realistic setup are consistent with those obtained with homogeneous phantoms, validating the method used in this study \cite{erdemlast, 10133485}. 

As the outer encapsulation strongly affects both the robustness and radiation performance of ingestible antennas, this study also investigates the effect of shell thickness on the gain and radiation efficiency ($\eta$) of these antennas at 434~MHz and reports $\eta$ values at operating frequencies allocated for medical communications. Additionally, the paper compares the $\eta$ results with the theoretical maximum achievable radiation efficiency~\cite{optimal}. Compared to the previous work~\cite{erdemlast, 10133485, eucap2022}, this paper is the first to examine the performance of different types of ingestible antennas in terms of their operational parameters for multiplexed gastrointestinal (GI) sensing and communication. The results presented provide essential guidelines for engineers designing multiplexed biosensing antennas for in-body applications.

This paper is organized as follows. Section \ref{sec:antennas} describes the antenna models used in the work and summarizes the process for prototyping and the preparation of the measurement setup. Section \ref{sec:sensoranalysis} details the parameters examined for comparison and presents the suitability analysis of the antennas as biosensors. Section \ref{sec:effgain} shares a further analysis of the gain and radiation efficiency of the antennas for different capsule shell thicknesses and different in-body frequencies. Finally, conclusions are drawn in Section \ref{sec:conc}.

\section{Antenna Models and Measurement Setup}
\label{sec:antennas}
\subsection{Antenna Models and Numerical Optimization}
\label{subsec:models}
\begin{figure}
\centering
\includegraphics[width=\columnwidth]{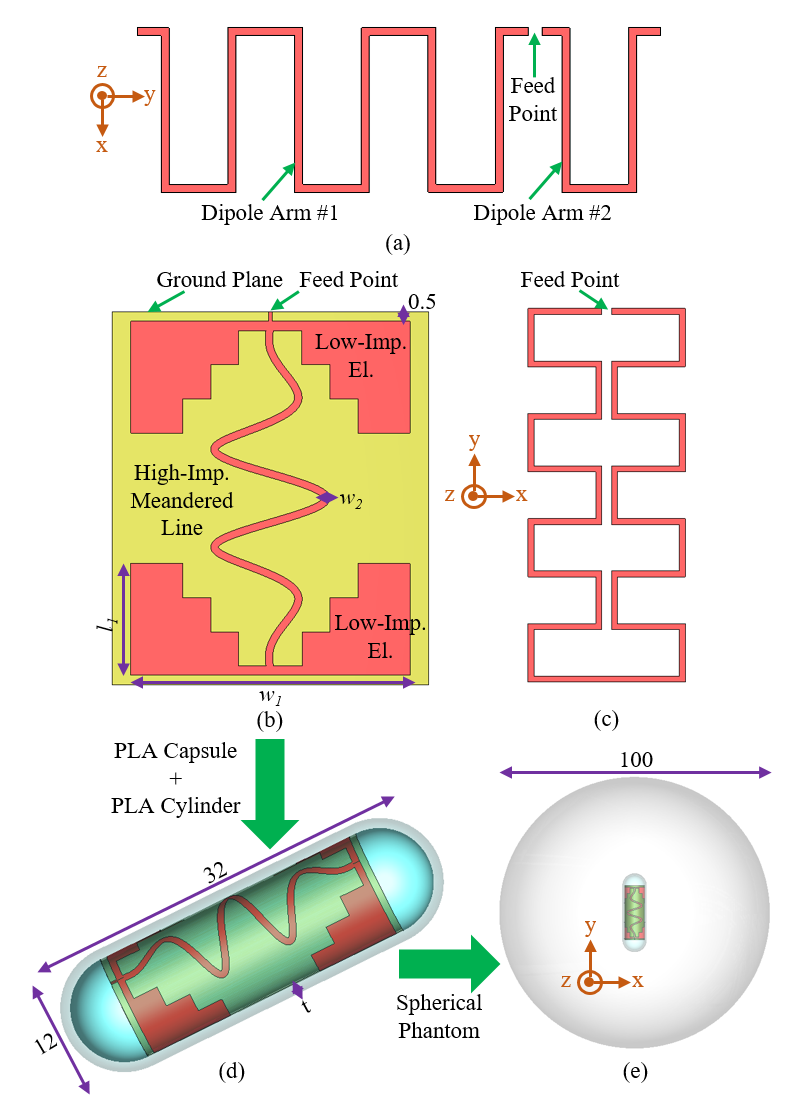}
\caption{Model of the antennas, the capsule, and the phantom used in the study (units: mm). (a) Meandered dipole antenna. (b) Patch antenna with two identical low-impedance radiating elements connected with a high-impedance meandered microstrip line. (c) Meandered loop antenna. (d) Patch antenna conforming to the inner surface of the PLA capsule filled with a PLA fixing cylinder. (e) Capsule placed in the middle of the spherical homogeneous phantom for EM simulations.}
\label{fig:antennas}
\end{figure}

Our analysis strategy is based on three common types of antennas [see Fig. \ref{fig:antennas}(a)--(c)]: the meandered dipole and the meandered loop antenna, which are presented in\cite{eucap2022}, and the patch antenna comprising of two identical low-impedance radiating elements connected with a high-impedance meandered microstrip line, which is reported in~\cite{denyspatchpaper}. 
The antennas are designed on a 100-\textmu m-thick Rogers CLTE-MW substrate ($\varepsilon_r$ = 2.97) \cite{rogers} and conform  to the inner surface of 32~mm $\times$ \diameter12~mm biocompatible polylactic-acid capsules (PLA, $\varepsilon_r = 2.7$, $\tan\delta = 0.003$) having a shell thickness $t$, as seen in Fig.~\ref{fig:antennas}(d). The capsules are  filled with PLA fixing cylinders to keep the antennas in their cylindrical shape inside the capsules. For the simulations, the capsules are placed in the middle of a spherical homogeneous phantom (\diameter100~mm) as shown in Fig. \ref{fig:antennas}(e). 

\begin{table}
\setlength{\tabcolsep}{1.5pt}
\renewcommand{\arraystretch}{1.5}
\caption{EM Properties of the GI Tissues and Measured Values for the Tissue-Mimicking Liquids at 434 MHz}
\label{tab:emproperties}
\centering
\begin{tabular}{c S[table-format=2.1] S[table-format=2.1] S[table-format=1.1]| S[table-format=1.2] S[table-format=2.2] S[table-format=1.1]}
\hline
& \multicolumn{3}{c|}{\pmb{$\varepsilon_r$}} & \multicolumn{3}{c}{\pmb{$\sigma$} \textbf{(S/m)}}\\
& \textbf{Real} & \textbf{Meas.}& \textbf{Err. (\%)}& \textbf{Real} & \textbf{Meas.}& \textbf{Err. (\%)}\\
\hline
\textbf{Stomach (ST)}&67.2 &67.1  &0.2 & 1.01 &1.02 &0.2 \\

\textbf{Small Intestine (SI)}&65.3 &64.1  &1.8 & 1.92 &1.91 &0.5 \\

\textbf{Large Intestine (LI)}&62 &62.9  &1.5 & 0.87& 0.88&1.1 \\

\hline
\end{tabular}
\end{table}

For the analysis presented in Section \ref{sec:sensoranalysis}, the antennas are optimized to operate in the 433 MHz ISM band for shell thicknesses of $t = 0.2$~mm, $t = 0.4$~mm, and $t = 0.6$~mm using CST Studio Suite (Dassault Systèmes Simulia Corp.). These thickness values fall within the range used in commercial devices and are selected to demonstrate the strong dependency of the examined parameters on shell thickness. This approach provides further insights into finding the trade-off between the sensitivity of the antenna to surrounding tissues and its radiation performance. The optimized values for this study of the parameterized dimensions of the antennas (with the parametrization provided in \cite{eucap2022} and \cite{denyspatchpaper}) are tabulated in Table \ref{table:optdimen}. The feed gap of all dipole antennas and the loop antenna for $t = 0.2$~mm equals 0.25 mm, whereas it is 1 mm for the loop antenna for $t = 0.4$~mm and $t = 0.6$~mm. For the patch antenna, the ground plane is 24~mm $\times$ 20~mm for $t = 0.2$~mm and 32~mm $\times$ 20~mm for $t = 0.4$~mm and $t = 0.6$~mm. Furthermore, the meandered microstrip line connecting the radiating elements is implemented as a cubic spline as presented in \cite{denyspatchpaper} in detail. Along the $x$-coordinate, the points of the spline are defined as [0, $-w_{s}$, $2w_{s}$, $-2w_{s}$, $2w_{s}$, $-w_{s}$, 0] for $t = 0.2$~mm, [0, $-1.3w_{s}$, $2.1w_{s}$, $-2.5w_{s}$, $2.1w_{s}$, $-1.3w_{s}$, 0] for $t = 0.4$~mm, and [0, $-1.3w_{s}$, $2.1w_{s}$, $-2.8w_{s}$, $2.1w_{s}$, $-1.3w_{s}$, 0] for $t = 0.6$~mm. All the dimensions are optimized in a way that the average of the maximum and minimum center frequency ($f_{c}$) observed in different GI tissues (see Table \ref{tab:emproperties}) falls between 433.5--434.5~MHz. For example, the optimized dipole antenna for $t = 0.2$~mm has the minimum $f_{c}$ as 425.7~MHz in the SI, whereas it has the maximum $f_{c}$ as 441.4~MHz in the LI (see Table \ref{table:results1}, $f_{c1}$ and $f_{c2}$), and the average of these two values (433.55~MHz) falls between 433.5--434.5 MHz. This optimization is preferred in order to have a standardized optimization in the GI tract for all antennas, ensuring a consistent comparison.

As for the analysis performed in Section \ref{sec:effgain}, first, the shell thickness is varied in the interval [0.05, 1]~mm in increments of 0.05~mm to examine the effect of shell thickness on the gain and $\eta$ of the antennas. Second, the antennas are optimized for $t = 0.2$~mm to operate at different frequencies allocated for in-body communications (403, 434, 868, 915, 1400, and 2450~MHz) \cite{6161600}. At these frequencies, their gain and $\eta$ values are reported, and the $\eta$ values are compared with the theoretical maximum achievable $\eta$  for the GI tract that are derived using the methodology explained in \cite{optimal}.

\begin{table}
\setlength{\tabcolsep}{4pt}
\renewcommand{\arraystretch}{1.2}
\caption{Optimized Values of the Parameterized Dimensions of the Antennas for $t = 0.2$ \si{\milli\meter}, $t = 0.4$ \si{\milli\meter}, and $t = 0.6$ \si{\milli\meter}}
\label{table:optdimen}
\centering
\begin{tabular}{c c c c c c c }
\hline
 &\pmb{$t$} &\textbf{Length} &\textbf{Width}&\textbf{Trace Width} &\textbf{Feed Offset} &\textbf{N}\\
&\textbf{(mm)}&\textbf{(mm)}&\textbf{(mm)}&\textbf{(mm)}&\textbf{(mm)}& \\
\hline
\multirow{3}{*}{\textbf{Dipole}}&0.2 & \multirow{3}{*}{\textcolor{white}{0}1.9}&13.7&\multirow{3}{*}{0.25}&\multirow{3}{*}{6.6} &\multirow{3}{*}{6}\\
&0.4 & &16.2&& &\\
&0.6 & &17.8&& &\\
\hline
\multirow{3}{*}{\textbf{Loop}}& 0.2 &\textcolor{white}{0}8.4& \multirow{3}{*}{\textcolor{white}{0}1\textcolor{white}{.0}}&0.25&\multirow{3}{*}{*}&\multirow{3}{*}{13}\\
&0.4 &10.8 &&0.2\textcolor{white}{0}& &\\
&0.6 &12.3 &&0.25& &\\
\hline
\end{tabular}
\begin{tabular}{c S[table-format=1.1] S[table-format=1.1] S[table-format=1.1] S[table-format=2.1] S[table-format=1.1]S[table-format=1.1] S[table-format=1.1]}
  &\pmb{$t$}&\pmb{$l_1$}&\pmb{$l_c$}&\pmb{$w_1$}&\pmb{$w_2$}&\pmb{$w_c$}&\pmb{$w_s$}\\
 &\textbf{(mm)}&\textbf{(mm)}&\textbf{(mm)}&\textbf{(mm)}&\textbf{(mm)}&\textbf{(mm)}&\textbf{(mm)}\\
\hline
\multirow{3}{*}{\textbf{Patch}}&0.2 &8.3& 2.6&21.4&0.8&3&4\\
&0.4 &8&2.5&31.1&1.4&4.4&5.5\\
&0.6 &8.1&2.5&31&1.4&4.1&5.5\\
\hline
\end{tabular}
\end{table}

\subsection{Prototyping and Measurement Setup}
\label{subsec:meassetup}
\begin{figure}
\centering
\includegraphics[width=\columnwidth]{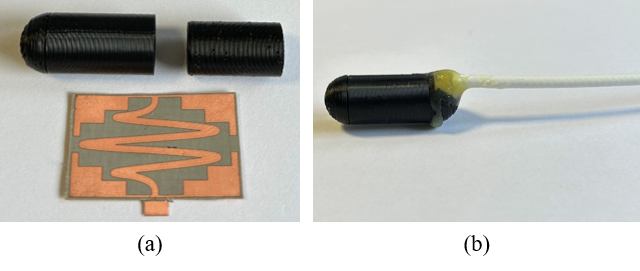}
\caption{Prototype used in the measurements. (a) Fabricated patch antenna, capsule, and fixing cylinder. (b) Final prototype with the coaxial cable and epoxy sealing.}
\label{fig:proto}
\end{figure}

 The measurements are conducted for $t = 0.4$~mm with the patch antenna. The optimized patch is fabricated using laser ablation technique, and the capsules and the fixing cylinders are fabricated with 3D-printing, as shown in Fig. \ref{fig:proto}(a). The fabricated antenna is soldered to a coaxial cable and put inside the capsule along with the fixing cylinder. Finally, the side of the capsule is sealed with Araldite 2012 epoxy resin. The final prototype can be seen in Fig. \ref{fig:proto}(b).

The measurements are conducted in liquids mimicking the EM properties of the GI tissues at 434 MHz. The tissue-mimicking liquids are prepared using water, sugar, and salt. The amount of each ingredient and the method of preparation can be found in \cite{erdemlast} and in \cite{liquidspaper}, respectively. The EM properties of the prepared liquids are measured using a SPEAG DAK-12 probe \cite{probe} and tabulated in Table \ref{tab:emproperties} along with the corresponding percentage errors between the real and measured values. Note that the maximum error equals 1.8\% for the relative permittivity and  1.1\% for the conductivity.

\section{Sensor Suitability Analysis and Comparison}
\label{sec:sensoranalysis}
\subsection{Definitions of Metrics}
To assess the suitability of an ingestible antenna as a biosensor for GI tissues, it is essential to examine several performance metrics. Firstly, the antenna must have sufficient sensing capability to reliably distinguish between different GI tissues. Secondly, it must exhibit robust impedance matching throughout the GI tract despite variations in the EM properties of the tissues. Finally, the antenna needs sufficient radiation efficiency to effectively transmit collected data through the lossy tissues to external devices. To analyze and compare the performance of the antennas based on these three metrics, the following antenna parameters are examined.

\subsubsection{Changes in Phase Values in Different Tissues}
As the antenna travels through the GI tract, it experiences detuning in its impedance response due to the varying EM properties of the GI tissues, as explained in Section~\ref{sec:introduction}. This detuning leads to changes in the reflection coefficient, affecting both its magnitude and phase. A previous proof-of-concept study by the authors demonstrated that phase changes in different GI tissues are sufficient to locate the capsule within the GI tract~ \cite{erdemlast}. Therefore, this study uses phase changes as the sensing parameter, where a change in phase indicates a change in the surrounding environment. Note that the capsule travels from ST to SI, and then from SI to LI; hence, the phase differences are examined in this sequence. 

\subsubsection{Frequency Interval for $|S_{11}|<$--10~dB in All Tissues}
The frequency interval in which the antennas are matched to 50~$\Omega$ in all GI tissues ($\Delta f_{i}$, Fig \ref{fig:resultst02}(a)) is examined to compare their robustness through the GI tract.

\subsubsection{Maximum Difference between the Center Frequency in Different Tissues}
The maximum difference between the center frequencies obtained in the tissues ($f_{c}$, Fig \ref{fig:resultst02}(a)) is also examined for the robustness of the antennas.  

\subsubsection{Gain}
The gain of the antennas is examined for their capability of transmitting the collected data to the outside of the body.

\subsubsection{Radiation Efficiency}
The radiation efficiency of the antennas is also examined separately of gain to compare their transmission capabilities decorrelated from directivity.

\subsection{Simulation Results}
\begin{figure*}
\centering
\includegraphics[width=2\columnwidth]{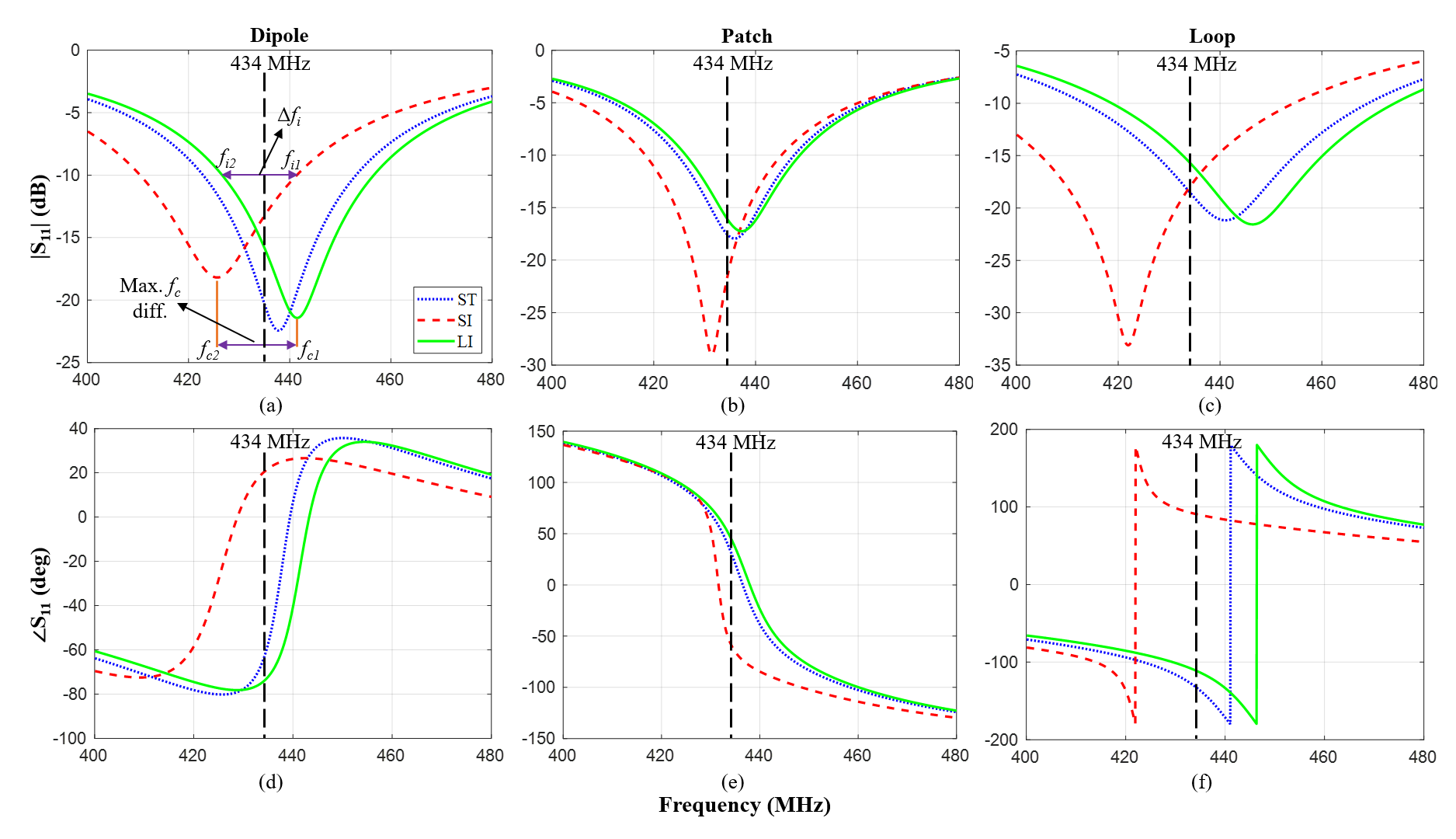}
\caption{Simulated magnitude and phase of the reflection coefficient of the antennas in three gastrointestinal tissues for $t = 0.2$~mm. Magnitude for the (a) dipole, (b) patch, and (c) loop. Phase for the (d) dipole, (e) patch, and (f) loop. Note that the measurements were conducted for $t = 0.4$~mm, and the measurement results are presented later in the paper (Fig.~\ref{fig:measresult}).}
\label{fig:resultst02}
\end{figure*}

\begin{figure*}
\centering
\includegraphics[width=2\columnwidth]{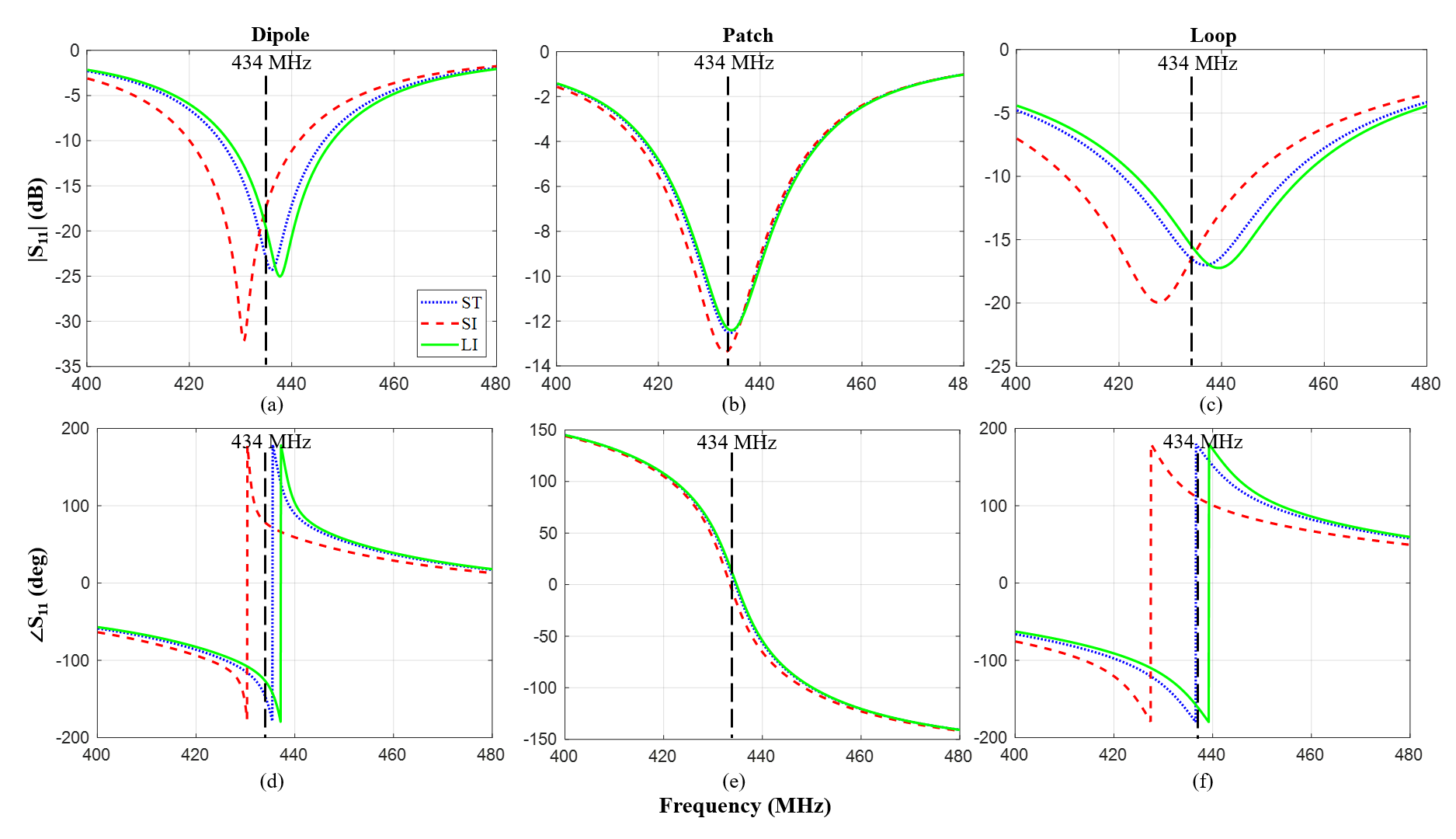}
\caption{Simulated magnitude and phase of the reflection coefficient of the antennas in three gastrointestinal tissues for $t = 0.6$~mm. Magnitude for the (a) dipole, (b) patch, and (c) loop. Phase for the (d) dipole, (e) patch, and (f) loop.}
\label{fig:resultst06}
\end{figure*}

Fig. \ref{fig:resultst02} and Fig. \ref{fig:resultst06} show the simulated magnitude and phase of the reflection coefficient of the antennas in three GI tissues for $t = 0.2$~mm and for $t = 0.6$~mm, respectively. Moreover, the simulated values of all examined parameters are tabulated in Table \ref{table:results1} for $t = 0.2$~mm and $t = 0.6$~mm and in Table \ref{table:results2} for $t = 0.4$~mm. It can be observed that for $t = 0.2$~mm, all antennas have sufficiently large differences in the phase to distinguish the tissues, with the loop having the largest differences. As the shell thickness increases, the differences decrease for the patch and the loop. The decrement is more dominant for the patch, with a minimum difference of 13.8\degree~for $t = 0.6$~mm, meaning that it is much harder to realize the sensing with the patch for this thickness. As for the dipole, the differences increase from $t = 0.2$~mm to $t = 0.4$~mm and decrease from $t = 0.4$~mm to $t = 0.6$~mm. The dipole and the loop have sufficiently large differences for all thicknesses, and the dipole has the greatest differences for $t = 0.4$~mm and $t = 0.6$~mm, making it the best option in terms of sensing capability for these thicknesses. 

It can also be seen that the loop has the greatest $\Delta f_{i}$ for all cases, which is consistent with~\cite{optimal} for this frequency range. With increasing shell thickness, $\Delta f_{i}$ decreases strongly for the patch and loop (49.7\% for the patch and 36.2\% for the loop from $t = 0.2$~mm to $t = 0.6$~mm), whereas the variations in $\Delta f_{i}$ are smaller than 2~MHz for the dipole. As for $f_{c}$, the patch has the lowest $f_{c}$ for all cases, with a maximum of 5.9~MHz at $t = 0.2$~mm, making it the most robust option against the EM changes in terms of the center frequency. Furthermore, $f_{c}$ decreases for all antennas with increasing thickness. 

\begin{figure}
\centering
\includegraphics[width=1\columnwidth]{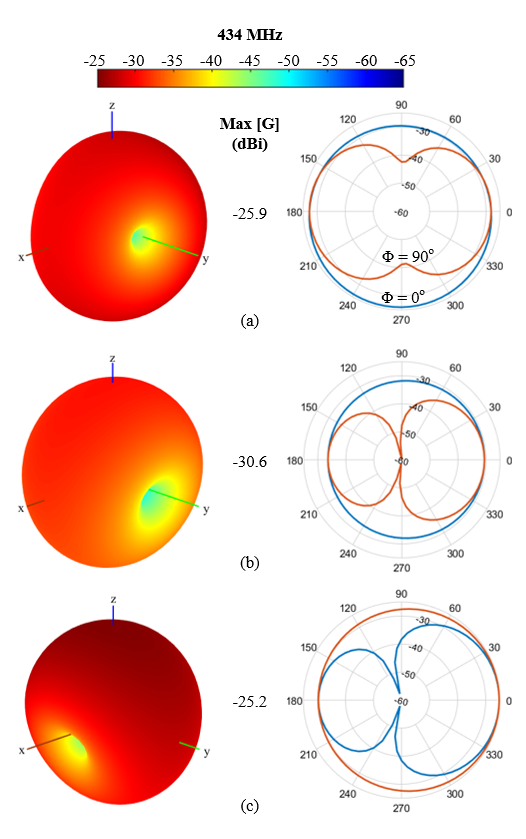}
\caption{Simulated 3D and 2D radiation patterns at 434 MHz  for the (a) dipole, (b) patch, and (c) loop in the LI for $t = 0.4$~mm (units: dBi for gain, degree for angle). The 2D patterns are given for $\Phi$ = 0$\degree$ (blue) and $\Phi$ = 90$\degree$ (orange) over the complete range of $\theta$ in spherical coordinates.}
\label{fig:radpat}
\end{figure}

As for the transmission capability, it can be seen from Fig. \ref{fig:radpat} that all antennas have a quasi-omnidrectional radiation pattern. Moreover, it can be observed in Table \ref{table:results1} and Table \ref{table:results2} that the loop has the largest gain and $\eta$ for all cases. The loop is closely followed by the dipole, with a maximum difference of 1.7~dBi for the gain and 1.1~dB for $\eta$ (both in SI, for $t = 0.4$~mm). Lastly, increasing the thickness results in an increment in the gain and $\eta$ for the dipole and loop, whereas it leads to a decrement for the patch, which will be examined more in detail in the next section.

In conclusion, these three types of antennas exhibit favorable characteristics for most of the cases, indicating their potential suitability as biosensors in the GI tract. Notably, the loop antenna emerges as the most attractive option, characterized by substantial phase differences and maximum values of $\Delta f_{i}$, gain, and $\eta$ for all cases. Moreover, these results also emphasize that it is essential for ingestible antenna designers to study the significant influence of shell thickness on antenna parameters. Finally, despite the comparatively less effective parameters of the patch in the context of biosensing in the GI tract in most of the cases, it must be stated that its ground plane serves as a protective measure against potential detuning resulting from other internal components within the capsule, thereby providing an advantage in EM compatibility.

\begin{table}
\setlength{\tabcolsep}{1pt}
\renewcommand{\arraystretch}{1.3}
\caption{Results for the Simulated Parameters of the Antennas for $t = 0.2$ \si{\milli\meter} and $t = 0.6$ \si{\milli\meter}}
\label{table:results1}
\centering
\begin{tabular}{c c| S[table-format=-2.1] S[table-format=-2.1] S[table-format=-3.1]| S[table-format=-3.1] S[table-format=-3.1] S[table-format=-3.1]}
\hline
&&\multicolumn{3}{c|}{\pmb{$t$}\textbf{ = 0.2~mm}}&\multicolumn{3}{c}{\pmb{$t$}\textbf{ = 0.6~mm}}\\
&&\textbf{Dipole} &\textbf{Patch} &\textbf{Loop} &\textbf{Dipole} &\textbf{Patch} &\textbf{Loop}\\
\hline
\multirow{2}{*}{\textbf{Phase}}&ST&-64.5  &33.4 &-131.5&-146.8&6.4&-158.5\\

\multirow{2}{*}{\textbf{(434 MHz, deg)}}&SI&20.1  &-56.9 & 91.1&78.2&-7.4& 124\\

&LI&-74.4 &46.6 &-110.5&-126.7&10.1&-139.3\\
\hline
\textbf{Phase Diff.}&ST to SI &84.6 &90.3&137.4&135&13.8&78  \\

\textbf{(434 MHz, deg)}&SI to LI&94.5 &103.5 &158.4&155.1&17.5&96.7\\
\hline
\multirow{2}{*}{\pmb{$\Delta f_{i}$}}&$f_{i1}$&441.4 &445&454.4&441.4&438.8&446\\

\multirow{2}{*}{\textbf{(MHz)}}&$f_{i2}$&426.7  &426.3&418.8&427.7&429.4&423.3  \\

&$\Delta f_{i}~(f_{i1}$ - $f_{i2})$&14.7 &18.7&35.6&13.7&9.4&22.7\\
\hline
\textbf{Max.} \pmb{$f_{c}$}&$f_{c1}$&441.4 &437.3&446.4&437.7&434.4&439.4\\

\textbf{Diff.}&$f_{c2}$&425.7  &431.4& 422&430.6&433&427.7\\

\textbf{(MHz)}&$f_{c1}$ - $f_{c2}$&15.7 &5.9 &24.4&7.1&1.4&11.7\\
\hline

\multirow{2}{*}{\textbf{Gain}}&ST&-27.9 &-31.2&-27.1&-25.8&-31.9&-25.3\\

\multirow{2}{*}{\textbf{(434 MHz, dBi)}}&SI&-35.1  & -36.8&-33.9&-32.4&-38.1&-31.7\\

&LI&-27.4 &-30.5&-26.6&-25.2&-31.6&-24.8\\
\hline
\multirow{2}{*}{\textbf{Rad. Eff.}}&ST&-26&-28.3&-25.9&-24.9&-30.2&-24.6 \\

\multirow{2}{*}{\textbf{(434 MHz, dB)}}&SI&-33.3  &-34.3 & -32.9&-31.8&-36.6&-31.2\\

&LI&-25.5 &-27.7&-25.4&-24.4&-29.9&-24.2\\
\hline

\end{tabular}
\end{table}

\begin{table} [htbp]
\setlength{\tabcolsep}{1.5pt}
\renewcommand{\arraystretch}{1.3}
\caption{Results for the Simulated and Measured Parameters of the Antennas for $t = 0.4$ \si{\milli\meter}}
\label{table:results2}
\centering
\begin{tabular}{c c| S[table-format=-3.1] S[table-format=-3.1] S[table-format=-3.1]| S[table-format=-3.1]}
\hline
&&\multicolumn{3}{c|}{\textbf{Sim.}}&\textbf{Meas.}\\
& &\textbf{Dipole} &\textbf{Patch} &\textbf{Loop}&\textbf{Patch}\\
\hline
\multirow{2}{*}{\textbf{Phase}}&ST&-104.9&21.9&-135.6&-103.1\\

\multirow{2}{*}{\textbf{(434 MHz, deg)}}&SI&43.3&-10.2&98.2&-119.3 \\

&LI&-104.5&29.6&-113.8&-102.5 \\
\hline
\textbf{Phase Diff.}&ST to SI &148.2&32.1&126.8&16.2  \\

\textbf{(434 MHz, deg)}&SI to LI&147.8&39.8&148&16.8\\
\hline
\multirow{2}{*}{\pmb{$\Delta f_{i}$}}&$f_{i1}$&442.2&441.4&448.5&441.7\\

\multirow{2}{*}{\textbf{(MHz)}}&$f_{i2}$&426.6&428.5&422.1&429.3  \\

&$\Delta f_{i}~(f_{i1}$ - $f_{i2})$&15.6&12.9&26.4&12.4\\
\hline
\textbf{Max.} \pmb{$f_{c}$}&$f_{c1}$&438.5&435.6&441.8&435.7\\

\textbf{Diff.}&$f_{c2}$&429.6&433.3&426.5&434.3\\

\textbf{(MHz)}&$f_{c1}$ - $f_{c2}$&8.9&2.3&15.3&1.4\\
\hline

\multirow{2}{*}{\textbf{Gain}}&ST&-26.9& -30.8&-25.6&\\

\multirow{2}{*}{\textbf{(434 MHz, dBi)}}&SI&-33.9&-36.9&-32.2&*\\

&LI&-25.9&-30.6&-25.2&\\
\hline
\multirow{2}{*}{\textbf{Rad. Eff.}}&ST&-25.5&-28.6&-24.8& \\

\multirow{2}{*}{\textbf{(434 MHz, dB)}}&SI&-32.5&-35&-31.4&*\\

&LI&-24.6&-28.4 &-24.4&\\
\hline
\end{tabular}
\end{table}

\subsection{Measurement Results}
\begin{figure}
\centering
\includegraphics[width=\columnwidth]{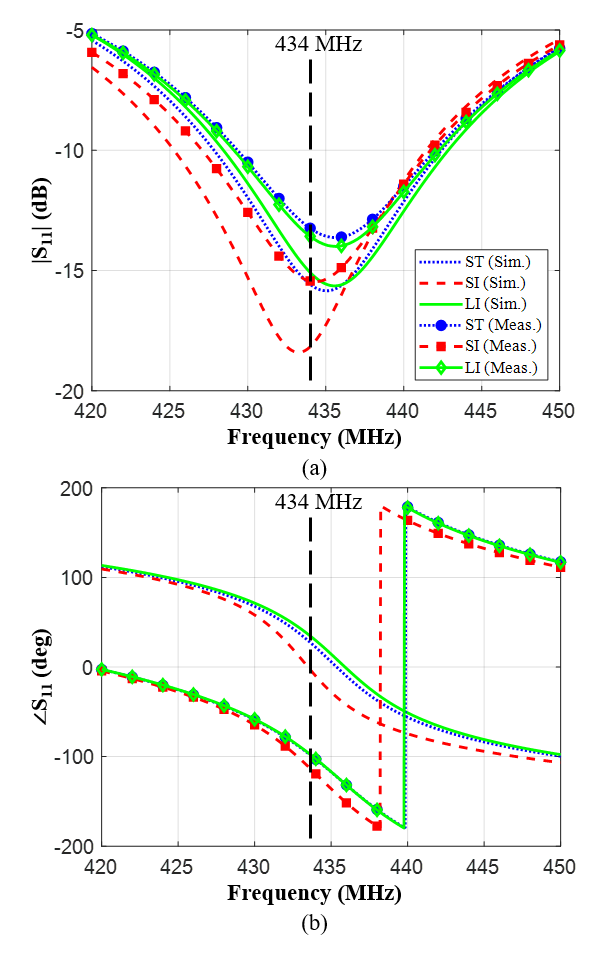}
\caption{Measurement results for the (a) magnitude and (b) phase of the reflection coefficient of the patch antenna for $t$ = 0.4~mm along with the simulation results for the same case. The two sub-figures share the same legend.}
\label{fig:measresult}
\end{figure}

Fig. \ref{fig:measresult}(a) and Fig. \ref{fig:measresult}(b) show the measured magnitude and phase of the reflection coefficient of the patch antenna, respectively, in three tissue-mimicking liquids for $t = 0.4$~mm along with the simulation results for the same case. The measured values of the parameters are tabulated in Table \ref{table:results2}. As can be seen from Fig. \ref{fig:measresult}(a), the measurements closely follow the simulations for the magnitude response. The measured values of $\Delta f_{i}$ and the maximum $f_{c}$ difference are in good agreement with the simulated ones, with only 0.5~MHz decrement for $f_{i}$ and 0.9~MHz decrement for the maximum $f_{c}$ difference. As for the phase presented in Fig. \ref{fig:measresult}(b), it can be observed that there is a shift in the values between the simulated and measured responses. This shift is around 115$\degree$ at 420~MHz [where the graphs start in Fig. \ref{fig:measresult}(b)] for all tissues, and it results from the coaxial cable used to connect the antenna to the Vector Network Analyzer (VNA). 

Table \ref{table:results2} shows that the phase differences observed in the tissues have diminished in the measurements. This is likely due to inaccuracy in the 3D-printing process, resulting in a shell thickness greater than 0.4 mm ($t = 0.43$ mm). This conclusion is also supported by the fact that $f_{i}$ and the maximum $f_{c}$ difference have slightly diminished in the measurements as well. Nevertheless, both the measured magnitude and phase responses exhibit similar tendencies to the simulated ones, demonstrating a sufficiently good agreement and thus validating the results presented in this study. The measurements were conducted only for $S_{11}$, as it is the parameter used for GI sensing, which is the main focus of this paper. The gain and $\eta$ values are provided through simulations to compare the transmission capabilities of the investigated antennas, with their measurements left out of the scope of this paper.

\section{Gain and Radiation Efficiency Analysis}
\label{sec:effgain}
In this section, a further analysis for the effect of the shell thickness on the gain and $\eta$ of the antennas at 434~MHz is presented. Moreover, $\eta$ of the antennas is examined at different frequencies allocated for in-body communications, and the values are reported along with a comparison with the theoretical maximum achievable $\eta$ derived for the GI tract. All the analysis in this section is conducted in phantoms having time-averaged EM properties of the GI tract (i.e. $\varepsilon_r$ = 63, $\sigma$ = 1.02~S/m at 434~MHz). The detailed fabrication procedure of this phantom can be found in \cite{eucap2022}. The dimensions of the phantom and the capsule are the same as in Fig. \ref{fig:antennas}.
\subsection{Effect of Shell Thickness}
In order to observe the effect of shell thickness on the gain and $\eta$ of the antennas, the shell thickness is varied from 0.05~mm to 1~mm with increments of 0.05~mm. When the shell thickness is increased,  a larger portion of the near-fields will be confined in the shell, which is expected to increase the gain and $\eta$ as the shell material is less lossy than the tissue. On the other hand, increasing the shell thickness results in less dielectric loading from the tissue, which increases the dimensions of the antennas and hence the extend of their near-field into the lossy tissue, causing more near-field losses. Thus, a trade-off exists in increasing the shell thickness, which is investigated at 434~MHz in this part of the study.

\begin{figure}
\centering
\includegraphics[width=0.85\columnwidth]{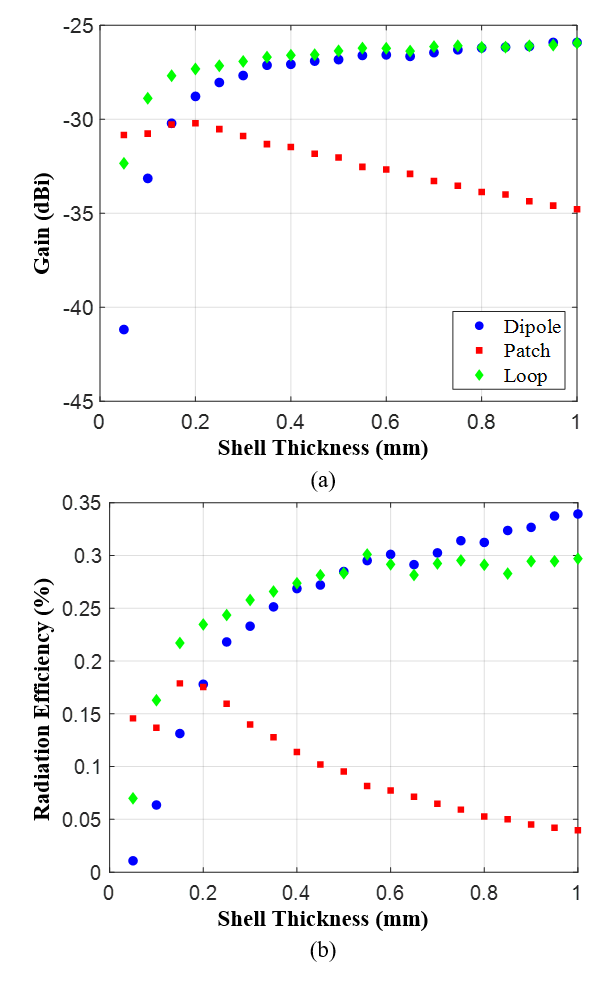}
\caption{Effect of the shell thickness on the (a) gain and (b) radiation efficiency ($\eta$)  of the antennas for $t \in $ [0.05, 1]~mm at 434~MHz. The two sub-figures share the same legend.}
\label{fig:shellthickres}
\end{figure}

Fig. \ref{fig:shellthickres}(a) and Fig. \ref{fig:shellthickres}(b) show the gain and $\eta$ of the antennas at 434 MHz with increasing shell thickness, respectively, and Table \ref{tab:gaineffvalues} tabulates the maximum and minimum values of these parameters along with the $t$ values at which the maximum or minimum is observed. As can be seen, the two graphs have similar tendencies for all antennas. For the dipole and loop, increasing the shell thickness results in increased gain and $\eta$. This effect is particularly substantial at thinner thicknesses (until $t = 0.2$~mm), with the rate of increment diminishing as thickness increases. This result demonstrates that the long meandering of these antennas along the surface fosters strong coupling of radiated fields with the surrounding tissue. This coupling decreases with increased thickness, as more and more fields are isolated from the tissue, leading to increased gain and $\eta$ in consequence.  Hence, this result signifies the prevailing influence of field confinement within the shell over the loading effect of the tissue for the dipole and loop. Conversely, for the patch, increasing the thickness decreases the gain and $\eta$ (for gain after $t = 0.2$~mm and for efficiency after $t = 0.15$~mm). This outcome indicates that the patch derives greater benefits from dielectric loading compared to other antennas. This inference is also supported by the fact that the patch has a greater gain and $\eta$ compared to the other antennas for $t = 0.05$~mm, when the dielectric loading is maximally effective, whereas the dipole and loop outperform the patch with increasing thickness.  

\begin{table}
\setlength{\tabcolsep}{1.5pt}
\renewcommand{\arraystretch}{1.5}
\caption{Maximum and Minimum Values of the Gain and Radiation Efficiency ($\eta$) of the Antennas at 434~MHz with Changing Shell Thickness}
\label{tab:gaineffvalues}
\centering
\begin{tabular}{c S[table-format=1.2] S[table-format=-2.3]| S[table-format=1.2] S[table-format=-2.3]| S[table-format=1.2] S[table-format=-2.3]}
\hline
& \multicolumn{2}{c|}{\textbf{Dipole}} & \multicolumn{2}{c|}{\textbf{Patch}} &\multicolumn{2}{c}{\textbf{Loop}} \\
& \pmb{$t$} \textbf{ (mm)} & \textbf{Value}& \pmb{$t$} \textbf{ (mm)} & \textbf{Value}&\pmb{$t$} \textbf{ (mm)} & \textbf{Value}\\
\hline
\textbf{Max. Gain (dBi)}&1 & -25.9 &0.2 &-30.2  &1 &-25.9 \\

\textbf{Min. Gain (dBi)}&0.05 & -41.2&1 &-34.8  &0.05 &-32.3\\
\hline
\textbf{Max. Rad. Eff. (\%)}&1 &0.339  &0.15 &0.179  &0.55 &0.301 \\

\textbf{Min. Rad. Eff. (\%)}&0.05 &0.011  &1 &0.04  &0.05 &0.07 \\

\hline
\end{tabular}
\end{table}

\subsection{Different Operating Frequencies}
In this section, the antenna efficiency values are analyzed at several frequencies allocated for medical communications depending on the region: 403, 434, 868, 915, 1400, and 2450~MHz \cite{6161600}. For this purpose, the three types of antennas are optimized to have $|S_{11}| < -10$~dB for $t = 0.2$~mm at each of these frequencies. Furthermore, theoretical bounds on maximum achievable $\eta$ are established for optimal electric and magnetic sources. The derivation of the theoretical bounds is performed using the methodology given in \cite{optimal} with the parameters $L$ = 20~mm, $R_{c}$ = 6~mm, $T$ = 0.2~mm, and $R_{p}$ = 50~mm, replicating the models implemented in this work. The phantom has the time-averaged EM properties of the GI tract that are calculated for the frequency interval [300--2500]~MHz using the approach explained in \cite{eucap2022} and the values in \cite{gabriel} and \cite{itis}. The theoretical bounds established for these source dimensions in the given standard phantom are compared with the obtained  $\eta$ values of the antennas. 

\begin{figure}
\centering
\includegraphics{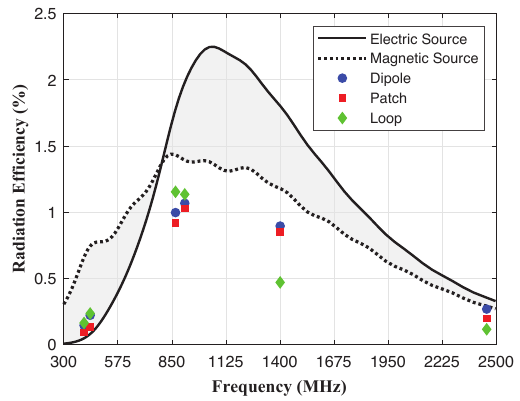}
\caption{Simulated $\eta$ of the antennas at different medical frequencies and the theoretical maximum achievable $\eta$ derived for the optimal electric and magnetic sources in the \diameter~= 100~mm homogeneous spherical phantom with time-averaged EM properties of the GI tract using the methodology presented in \cite{optimal}. Each antenna is optimized to have $|S_{11}| < -10$~dB for $t = 0.2$~mm at the investigated frequencies.}
\label{fig:graphfreq}
\end{figure}

Fig.~\ref{fig:graphfreq} shows the simulated $\eta$ of the antennas at the specified frequencies along with the derived theoretical bounds for the electric and magnetic sources. As observed, the $\eta$ values of the antennas follow a similar pattern to the optimal sources. Efficiencies of all antennas exceed the lower $\eta$ bounds of the theoretical electric source at 403 and 434~MHz (but remain below the higher $\eta$ bounds). This is because the proposed antennas benefit from dielectric loading and exhibit both electric and magnetic behavior, with one being more dominant depending on the antenna type.  The $\eta$ values fall below the bounds of both sources at higher frequencies. This is due to several factors affecting the operation of realistic antennas, such as dielectric loading by the tissues (which increases $\eta$) and ohmic~\cite{liska_maximum_2024} and dielectric losses (which decrease $\eta$). 

The dipole and patch antennas are most efficient at 915~MHz (1.068\% for the dipole and 1.028\% for the patch), whereas the loop antenna is most efficient at 868~MHz (1.154\%). Moreover, the loop outperforms the dipole and patch at lower frequencies, as predicted by the behavior of the optimal magnetic source,  consistent with previous findings~\cite{optimal}. It can also be seen that all antennas have lower $\eta$ values at higher frequencies. This decrease at higher frequencies can be attributed to increased attenuation in the tissues and the increased directivity of the antennas at these frequencies. Increased directivity indicates the presence of higher-order modes in the patterns, which increases power dissipation in the near-field \cite{5617242}.

\section{Conclusion}
The first part of this paper examined and compared the suitability of dipole, patch, and loop antennas as biosensors to distinguish GI tissues. The antennas were designed to conform to the inner surface of PLA capsules and operate in the 434 MHz ISM band. The results showed that all three types -- when properly optimized -- are suitable for GI sensing, with the loop antenna exhibiting marginally better performance in most cases. 

The second part of the paper investigated the effect of shell thickness on the gain and radiation efficiency of these antennas at 434 MHz, demonstrating that the patch antenna benefits more from the dielectric loading of the tissues than the other types. Additionally, this part compared the efficiencies of the antennas at various in-body frequencies with the theoretical maximum achievable efficiencies for the GI tract. The findings indicated that magnetic antennas outperform electric ones at lower frequencies in terms of efficiency, while electric antennas have greater efficiencies at higher frequencies, consistent with theoretical bounds derived in \cite{optimal}. The results presented in this paper provide groundwork data for designing multiplexed ingestible antennas for simultaneous sensing and data transmission.

\label{sec:conc}

\section*{Acknowledgement}
The authors would like to thank BodyCAP, without whom this research would have been impossible, and Christophe Guitton for the manufacturing of prototypes.

\bibliographystyle{IEEEtran}
\bibliography{References}
\end{document}